\documentclass[submission,copyright,creativecommons]{eptcs}
\providecommand{\event}{TFPIE 2018} 
\usepackage{amssymb}
\usepackage{breakurl}             
\usepackage{underscore}           
\usepackage[dvips]{graphicx}
\usepackage[dvips]{color}
\usepackage{alltt}

\newcommand{\ie}{\textit{i.e.}}
\newcommand{\eg}{\textit{e.g.}}
\newcommand{\wcm}{\texttt{w-c-m}}
\newcommand{\ccm}{\texttt{c-c-m}}

\definecolor{lightgray}{rgb}{0.83, 0.83, 0.83}
\definecolor{lightgreen}{rgb}{0.71, 0.93, 0.71}
\definecolor{purple}{rgb}{0.70, 0.50, 1.00}
\definecolor{red}{rgb}{1.00, 0.00, 0.00}

\title{Stepping OCaml}
\author{
Tsukino Furukawa \qquad\qquad Youyou Cong \qquad\qquad Kenichi Asai
\institute{Ochanomizu University\\ Tokyo, Japan}
\email{\{furukawa.tsukino, so.yuyu, asai\}@is.ocha.ac.jp}
}
\def\titlerunning{Stepping OCaml}
\def\authorrunning{T. Furukawa, Y. Cong \& K. Asai}
\begin{document}
\maketitle

\begin{abstract}
Steppers, which display all the reduction steps of a given program, are a novice-friendly tool for understanding program behavior.
Unfortunately, steppers are not as popular as they ought to be;
indeed, the tool is only available in the pedagogical languages of the DrRacket programming environment.

We present a stepper for a practical fragment of OCaml.
Similarly to the DrRacket stepper, we keep track of evaluation contexts in order to reconstruct the whole program at each reduction step.  The difference is that we support effectful constructs, such as exception handling and printing primitives, allowing the stepper to assist a wider range of users.  In this paper, we describe the implementation of the stepper, share the feedback from our students, and show an attempt at assessing the educational impact of our stepper.

\end{abstract}

\section{Introduction}
\label{sec:intro}

Programmers spend a considerable amount of time and effort on debugging.  In particular, novice programmers may find this process extremely painful, since existing debuggers are usually not friendly enough to beginners.  To use a debugger, we have to first learn what kinds of commands are available, and figure out which would be useful for the current purpose.  It would be even harder to use the command in a meaningful manner: for instance, to spot the source of an unexpected behavior, we must be able to find the right places to insert breakpoints, which requires some programming experience.

\begin{figure}
\begin{center}
\includegraphics[width=10cm]{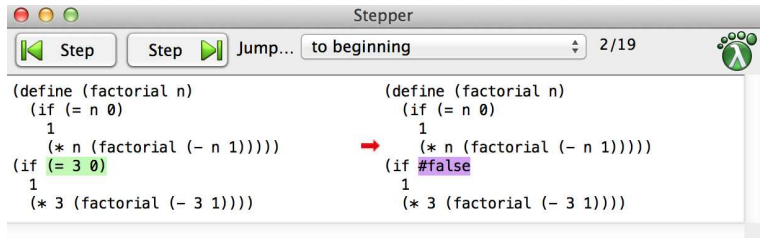}
\caption{Stepping factorial in DrRacket}
\label{figure:racketstep}
\end{center}
\end{figure}

Then, is there any debugging tool that is accessible to first-day programmers?  In our view, the algebraic stepper \cite{clements01} of DrRacket, a pedagogical programming environment for the Racket language, serves as such a tool.  The algebraic stepper is literally a stepping evaluator for DrRacket programs.  Figure \ref{figure:racketstep} illustrates how the stepper works.  In the ``before'' window (left), we see that the expression \texttt{(= 3 0)} is what we are going to reduce at the current step, i.e., the expression is a \emph{redex}.  In the ``after'' window (right), we find the redex is replaced by \texttt{false}, which is the result of reducing \texttt{(= 3 0)}.  

Note that there is a notable difference between the DrRacket stepper and the
stepping facility of Eclipse or gdb.  While Eclipse's debugger only shows
which line is being executed, the stepper tells us how the whole
program looks like at each step, by rewriting the input program
according to the reduction semantics of Racket.
Using the stepper would be easier for beginners, because all they need
to understand is how programs are rewritten as the execution proceeds,
just like how formulas are rewritten in mathematics.
Thus, we see that the DrRacket stepper should
serve as an excellent debugging tool for beginning programmers.

\begin{figure}
  \begin{center}
    \includegraphics[width=10cm]{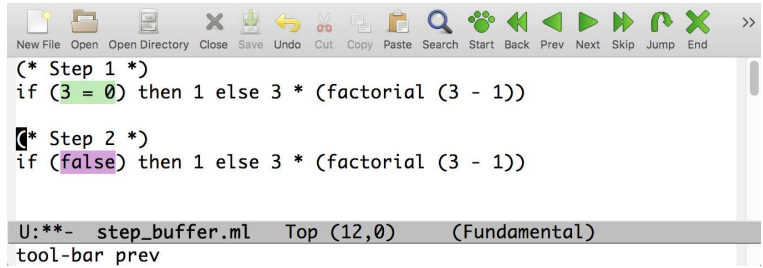}
    \caption{Stepping factorial in OCaml}
    \label{figure:ocamlfac}
  \end{center}
\end{figure}

Unfortunately, the DrRacket stepper is only available in the ``teaching languages''; it does not support the full Racket.  What this means is that we cannot step through programs that uses advanced features, including exception handling.  This is quite disappointing, since understanding how exceptions are handled is not easy for beginners.  

We present a stepper that supports a practical fragment of OCaml, covering most language constructs used in the ``Functional Programming'' course of Ochanomizu University.  The stepper looks exactly like the one provided in DrRacket, as shown in Figure \ref{figure:ocamlfac}.
We build the stepper by making modifications to a standard big-step interpreter.
The idea is to keep track of evaluation contexts and to reconstruct the whole program at each reduction step,
which are necessary for reconstructing the whole program when reducing redexes.
We also share our experience of using the stepper in the classroom, and attempt to evaluate the impact on students' understanding.

The reminder of this paper is organized as follows.
In Section~\ref{sec:imp}, we describe the implementation of a
mini stepper that supports exception handling.
It also shows the overview of the actually implemented stepper.
In Section~\ref{sec:exp}, we present how the stepper is used in our
classroom, as well as what students found about it in the recent courses.
Related work is discussed in Section~\ref{sec:related}, and the paper concludes
in Section~\ref{sec:conc}.

\section{Implementation}
\label{sec:imp}

This section presents the implementation of our stepper.  We step-evaluate programs in the following way.
First, we convert a given program into an abstract syntax tree using the built-in parser of OCaml.
Next, we pass the parsed program to a stepping interpreter,
which outputs the whole program at each reduction step.
The output programs are then processed by an Emacs Lisp program,
in such a way that the user can see the steps one by one.  Here we focus our attention to the stepping interpreter,
which plays the key role in the whole stepping system.

As the stepper is supposed to tell us how a program is evaluated, we have to make sure that it evaluates programs in the same order as the OCaml interpreter does.
For the presentation issue, here we restrict ourselves to a toy language,
consisting of lambda terms and a simplified version of the try-with construct.
The full language supported by the current stepper is presented in Section \ref{sec:imp:ocaml}.  

\subsection{Building an Interpreter}
\label{sec:imp:interpreter}

\begin{figure}
\begin{verbatim}
type e_t = 
         | Var of string               (* x *)
         | Fun of string * e_t         (* fun x -> e *)
         | App of e_t * e_t            (* e e *)
         | Try of e_t * string * e_t   (* try e with x -> e *)
         | Raise of e_t                (* raise e *)
\end{verbatim}
\caption{Syntax}
\label{figure:typee}
\end{figure}

\begin{figure}
\begin{verbatim}
(* exception holding the value of input program's exception *)
exception Error of e_t

(* evaluate expression *)
(* eval : e_t -> e_t *)
let rec eval expr = match expr with
  | Var (x) -> failwith ("unbound variable: " ^ x)
  | Fun (x, e) -> Fun (x, e)
  | App (e1, e2) ->
    begin
      let v2 = eval e2 in
      let v1 = eval e1 in
      match v1 with
      | Fun (x, e) ->
        let e' = subst e x v2 in   (* substitute v2 for x in e *)
        let v = eval e' in
        v
      | _ -> failwith "not a function"
    end
  | Try (e1, x, e2) ->
    begin
      try
        let v1 = eval e1 in
        v1
      with Error (v) ->
        let e2' = subst e2 x v in   (* substitute v for x in e2 *)
        eval e2'
    end
  | Raise (e) ->
    let v = eval e in
    raise (Error (v))

(* start evaluation *)
(* start : e_t -> e_t *)
let start e =
  try
    eval e
  with
    Error v -> Raise v
\end{verbatim}
\caption{Big-step interpreter}
\label{figure:interpreter}
\end{figure}

In Figures \ref{figure:typee} and \ref{figure:interpreter},
we define the object language as well as a big-step interpreter.
The \texttt{eval} function evaluates a given expression following OCaml's call-by-value, right-to-left strategy.
For instance, when given an application \texttt{e1 e2},
it first evaluates the argument \texttt{e2}, then evaluates the function \texttt{e1}.
Once the application has been turned into a redex, we perform $\beta$-reduction, and evaluate the post-reduction expression.
Note that, when the top-level expression is an executable, closed program, the input of the \texttt{eval} function cannot be a variable.  The reason is that we never touch a function's body before it receives an argument, and that $\beta$-reduction replaces lambda-bound variables with values.

Object-level exception handling is performed by the meta-level \texttt{try} and \texttt{raise} constructs.
Specifically, when evaluating \texttt{raise e},
we first evaluate \texttt{e} to some value \texttt{v},
and then raise a meta-level (OCaml) exception \texttt{Error v}.
If an exception \texttt{Error v} was raised during evaluation of \texttt{e1} in \texttt{try e1 with x -> e2}, the \texttt{eval} function ignores the rest of the computation in \texttt{e1}, and evaluates \texttt{e2} with \texttt{v} substituted for \texttt{x}.  This is exactly how OCaml's try-with construct works.  For convenience, we will hereafter call \texttt{e1} a \emph{tryee}; the intention is that \texttt{e1} is the expression being ``tried'' by the handler.   

The main function \texttt{start} calls \texttt{eval} in an exception handling context.  From the construction, we can see that any expression that has a \texttt{raise e} with no matching \texttt{try} clause will be evaluated to \texttt{raise v}.  For example, \texttt{2 + 3 + (raise 4) + 5} evaluates to \texttt{raise 4}.

\subsection{Turning the Interpreter into a Stepper}
\label{sec:imp:stepper}

As stated in Section \ref{sec:intro}, a stepper must display the whole program at each reduction step.  Consider the simple arithmetic expression \texttt{(1 + 2 * 3) + 4}.  When step-executing this expression, we want to see the following reduction sequence:
\[
\begin{array}{cl}
            & \mathtt{(1 + 2 * 3) + 4} \\
\rightarrow & \mathtt{(1 + 6) + 4} \\
\rightarrow & \mathtt{7 + 4} \\
\rightarrow & \mathtt{11}
\end{array}
\]

The interpreter in Figure \ref{figure:interpreter}, however, does not immediately give us these steps.  Suppose the \texttt{eval} function is evaluating the subexpression \texttt{2 * 3}.  We can display this subexpression using a printing function, but we do not have enough information to reconstruct the whole program.  What is missing here is the \emph{context} surrounding \texttt{2 * 3}, namely \texttt{(1 + [.])\ + 4} (where \texttt{[.]} denotes the hole of the context).  Hence, to implement a stepper, we need to keep track of every evaluation context we have traversed.  

In Figure \ref{figure:simpleplug}, we define context frames as algebraic data of type \texttt{frame\_t}.  Each frame represents evaluation of some subexpression: \eg, \texttt{CAppR (e1)} tells us that we are evaluating the argument part of an application, whose function part is \texttt{e1}.  Evaluation contexts are defined as lists of these frames (spoiler alert: this does not work for exceptions).  We then define the \texttt{plug} function, which reconstructs a program by wrapping the expression \texttt{expr} with context frames in \texttt{ctxt}. 

\begin{figure}
\begin{verbatim}
(* context frames *)
type frame_t = 
             | CAppR of e_t           (* e [.] *)
             | CAppL of e_t           (* [.] v  *)
             | CTry of string * e_t   (* try [.] with x -> e *)
             | CRaise                 (* raise [.] *)

(* evaluation contexts *)
type c_t = frame_t list

(* reconstruct the whole program *)
(* plug : e_t -> c_t -> e_t *)
let rec plug expr ctxt = match ctxt with
  | [] -> expr
  | CAppR (e1) :: rest -> plug (App (e1, expr)) rest
  | CAppL (e2) :: rest -> plug (App (expr, e2)) rest
  | CTry (x, e2) :: rest -> plug (Try (expr, x, e2)) rest
  | CRaise :: rest -> plug (Raise expr) rest
\end{verbatim}
\caption{Contexts and reconstruction function; first attempt}
\label{figure:simpleplug}
\end{figure}

Now, if we let the evaluation function receive an additional argument representing the context, we should be able to display all the steps of the arithmetic expression \texttt{(1 + 2 * 3) + 4}.  For instance, when evaluating the subexpression \texttt{2 * 3}, the extra argument will be a two-element list \texttt{[(1 + [.]);\ ([.]\ + 4)]}, and we can obtain the whole program using the \texttt{plug} function.

The resulting stepper is essentially the CK abstract machine
\cite{FF1986}, where the expression is the control string and the
evaluation context is the continuation.
Substitution is used to implement $\beta$-reduction.
We did not implement the abstract machine directly but augmented a
big-step interpreter, because we want to keep the correspondence between
big-step execution and small-step execution.
It enables us to skip evaluation of user-specified function application,
as we elaborate in Section~\ref{sec:imp:ocaml}.

Unfortunately, this na\"ive implementation does not work in the presence of exception handlers.  Consider \texttt{try (2 + 3 * (raise 4) + 5) with x -> x}.  When step-executing this expression, we expect to see the following steps:

\vspace{0.2cm}

\noindent \texttt{(* Step 0 *) try (\colorbox{lightgreen}{2 + (3 * (raise 4)) + 5}) with x -> x\\
  (* Step 1 *) try (\colorbox{purple}{raise 4}) with x -> x\\
  (* Step 1 *) \colorbox{lightgreen}{try (raise 4) with x -> x}\\
  (* Step 2 *) \colorbox{purple}{4}\\
}

\vspace{0.2cm}

\noindent The first reduction happens when the input to the stepping interpreter is \texttt{raise 4}.  However, observe that the highlighted redex is a bigger expression \texttt{(2 + 3 * (raise 4) + 5)}, because reduction of a \texttt{raise} construct discards the context \emph{within} the tryee.  Since context frames are collected in a single list, the second argument at this point will be \texttt{[(3 * [.]); (2 + [.]);\ ([.]\ + 5);\ (try [.]\ with x -> x)]}, \ie, it contains the context \emph{outside} the tryee.  This suggests that, when dealing with exception handlers, we have to distinguish between contexts inside and outside a tryee.\footnote{
The destination is not necessary,
if we want to support only exception handling.
We could simply search for the enclosing handler in the evaluation
context.
However, it requires a linear search through the evaluation context.
Furthermore, distinction is necessary if we want to implement
more general control operators,
such as shift and reset \cite{DF1990}.
}

\begin{figure}
\begin{verbatim}
(* frames *)
type frame_t = 
             | CAppR of e_t          (* e [.] *)
             | CAppL of e_t          (* [.] v *)
             | CRaise                (* raise [.] *)

(* try frame *)
type ctry_t = 
            | CHole                        (* [.] *)
            | CTry of string * e_t * c_t   (* try [.] with x -> e *)

(* evaluation context *)
and c_t = frame_t list * ctry_t

(* reconstruct tryee *)
(* plug_in_try : e_t -> frame_t list -> e_t *)
let rec plug_in_try expr ctxt = match ctxt with
  | [] -> expr
  | first :: rest -> match first with
    | CAppR (e1) -> plug_in_try (App (e1, expr)) rest
    | CAppL (e2) -> plug_in_try (App (expr, e2)) rest
    | CRaise -> plug_in_try (Raise (expr)) rest

(* reconstruct the whole program *)
(* plug : e_t -> c_t -> e_t *)
let rec plug expr (clist, tries) =
  let tryee = plug_in_try expr clist in
  match tries with
  | CHole -> tryee
  | CTry (x, e2, outer) -> plug (Try (tryee, x, e2)) outer
\end{verbatim}
  \caption{Contexts and reconstruction function; final version}
  \label{figure:typec}
\end{figure}

In Figure \ref{figure:typec}, we present a refined definition of evaluation contexts.  We see a new definition of context frames \texttt{frame\_t}, where \texttt{CTry} is missing.  When evaluating a program that uses try-with constructs, these frames are used to build a delimited context within a tryee.  We next find a separate datatype \texttt{ctry\_t}, which can be understood as meta contexts.  Then we define evaluation contexts as pairs of delimited and meta contexts.  As an example, when evaluating \texttt{raise 4} in the following expression:
\begin{verbatim}
0 + (try 1 + 2 * (try (3 + raise 4) - 5 with x -> x + 6) with y -> y)
\end{verbatim}
\noindent the current context looks like:
\begin{verbatim}
([(3 + [.]); ([.] - 5)],
  CTry ("x", x + 6,
    ([2 * [.]; 1 + [.]],
      CTry ("y", y,
        ([0 + [.]], CHole)))))
\end{verbatim}

The refined contexts allow us to first reconstruct the expression up to the tryee using the \texttt{frame\_t} contexts, and then build up the whole program using the \texttt{ctry\_t} contexts.  In our particular example, the stepper reconstructs \texttt{(3 + raise 4) - 5}, highlights it, and reconstructs the whole program.

To give the reader a better idea how context frames are accumulated, let us demonstrate the evaluation of an expression involving exception handling:

\begin{verbatim}
eval (2 * (try 3 + (raise 4) - 5 with x -> x + 6)) ([], CHole)
eval (try 3 + (raise 4) - 5 with x -> x + 6) ([2 * [.]], CHole)
eval (3 + (raise 4) - 5) ([], CTry ("x", x + 6, ([2 * [.]], CHole)))
eval 5 ([3 + (raise 4) - [.]], CTry ("x", x + 6, ([2 * [.]], CHole)))
eval (3 + (raise 4)) ([[.] - 5], CTry ("x", x + 6, ([2 * [.]], CHole)))
eval (raise 4) ([3 + [.]; [.] - 5], CTry ("x", x + 6, ([2 * [.]], CHole)))
eval 4 ([raise [.]; 3 + [.]; [.] - 5], CTry ("x", x + 6, ([2 * [.]], CHole)))
eval (4 + 6) ([2 * [.]], CHole)
\end{verbatim}

\noindent Observe that we discard the context within the tryee, namely \texttt{3 + (raise [.])\ - 5}, at the last step.

Now we present our stepping interpreter in Figure \ref{figure:stepper}.  The function extends the big-step interpreter in two ways (as shaded in the figure): (i) it receives an argument representing the evaluation context; and (ii) it outputs the current program every time reduction takes place.  

\begin{figure}
%

\begin{alltt}
(* stepping evaluator *)
(* eval : e_t -> c_t -> e_t *)
let rec eval expr \colorbox{lightgray}{ctxt} = match expr with       (* add an argument for context *)
  | Var (x) -> failwith ("unbound variable: " ^ x)
  | Lam (x, e) -> Lam (x, e)
  | App (e1, e2) ->
    begin
      let v2 = eval e2 \colorbox{lightgray}{(add ctxt (CAppR e1))} in           (* add context info *)
      let v1 = eval e1 \colorbox{lightgray}{(add ctxt (CAppL v2))} in           (* add context info *)
      match v1 with
      | Lam (x, e) ->
        let e' = subst e x v2 in
        \colorbox{lightgray}{memo (App (v1, v2)) e' ctxt;}                       (* output programs *)
        let v = eval e' \colorbox{lightgray}{ctxt} in                           (* add context info *)
        v
      | _ -> failwith "not a function"
    end
  | Try (e1, x, e2) ->
    begin
      try
        let v1 = eval e1 \colorbox{lightgray}{(add_try ctxt x e2)} in           (* add context info *)
        \colorbox{lightgray}{memo (Try (v1, x, e2)) v1 ctxt;}                    (* output programs *)
        v1
      with Error (v) ->
        let e2' = subst e2 x v in
        \colorbox{lightgray}{memo (Try (Raise v, x, e2)) e2' ctxt;}              (* output programs *)
        eval e2' \colorbox{lightgray}{ctxt}                                     (* add context info *)
    end
  | Raise (e0) ->
    let v = eval e0 \colorbox{lightgray}{(add ctxt CRaise)} in                  (* add context info *)
    \colorbox{lightgray}{begin match ctxt with                   }
\end{alltt}
\vspace{-18pt}
\begin{alltt}
    \colorbox{lightgray}{    | ([], _) -> ()                     }
\end{alltt}
\vspace{-18pt}
\begin{alltt}
    \colorbox{lightgray}{    | (clist, tries) ->                 }
\end{alltt}
\vspace{-18pt}
\begin{alltt}
    \colorbox{lightgray}{      memo (plug_in_try (Raise v) clist)}               (* output programs *)
\end{alltt}
\vspace{-18pt}
\begin{alltt}
    \colorbox{lightgray}{           (Raise v)                    }
\end{alltt}
\vspace{-18pt}
\begin{alltt}
    \colorbox{lightgray}{           ([], tries)                  }
\end{alltt}
\vspace{-18pt}
\begin{alltt}
    \colorbox{lightgray}{end;                                    }
    raise (Error (v))
\end{alltt}
\caption{Stepping evaluator}
\label{figure:stepper}
\end{figure}

\begin{figure}
\begin{alltt}
(* output programs *)
(* memo : e_t -> e_t -> c_t -> unit *)
let memo expr1 expr2 ctxt =
  print_exp (plug (green expr1) ctxt);
  print_exp (plug (purple expr2) ctxt)

(* start step-evaluation *)
(* start : e_t -> e_t *)
let start e =
  try
    eval e \colorbox{lightgray}{([], CHole)}    (* initial context *)
  with
    Error (v) -> (Raise v)
\end{alltt}
\caption{\texttt{memo} and main functions}
\label{figure:memo}
\end{figure}

Let us observe the application case.  As in the big-step interpreter, we first evaluate \texttt{e2}, and then \texttt{e1}.  When \texttt{e1} has reduced to a function, we know that the application is a $\beta$-redex.  In the standard interpreter, what we do is to perform the substitution \texttt{subst e x v2} and then evaluate the result.  In the stepper, on the other hand, we have an additional function call to the \texttt{memo} function defined in Figure \ref{figure:memo}.  This function receives three arguments: the redex we have just found, its reduct, and the current evaluation context.  When given these arguments, the \texttt{memo} function reconstructs and prints the pre- and post-reduction programs, using the \texttt{plug} and \texttt{print\_exp} functions\footnote{In the actual implementation, we annotate redexes and reducts using OCaml's \emph{attributes}.  Here, we write \texttt{green expr1} to mean \texttt{expr1[@stepper.redex]}, and similarly for \texttt{purple}.  When displaying the steps, the Emacs Lisp program uses the attributes information to appropriately highlight expressions.}.  After printing the programs, we continue evaluation as usual.

In the \texttt{eval} function, we find three more occurrences of \texttt{memo}, representing the following reduction rules:

\begin{itemize}
  \item \texttt{try v with x -> e} $\leadsto$ \texttt{v}
  \item \texttt{try raise v with x -> e2} $\leadsto$ \texttt{subst e2 x v}
  \item \texttt{...\ (raise v) ...} $\leadsto$ \texttt{raise v}
\end{itemize}

\noindent Note that, although the second reduction always happens right after the third one, we keep them as separate rules.  The reason is that we need the latter to reduce a raise construct with no matching try clause: \eg, \texttt{3 + (raise 4) - 5} $\leadsto$ \texttt{raise 4}.  Separating the two reductions also has an educational benefit: it clearly tells us that exception handling consists of two tasks: discarding the context and substituting the value.



\subsection{The Actual Stepper}
\label{sec:imp:ocaml}

\begin{figure}
  \includegraphics[width=14cm]{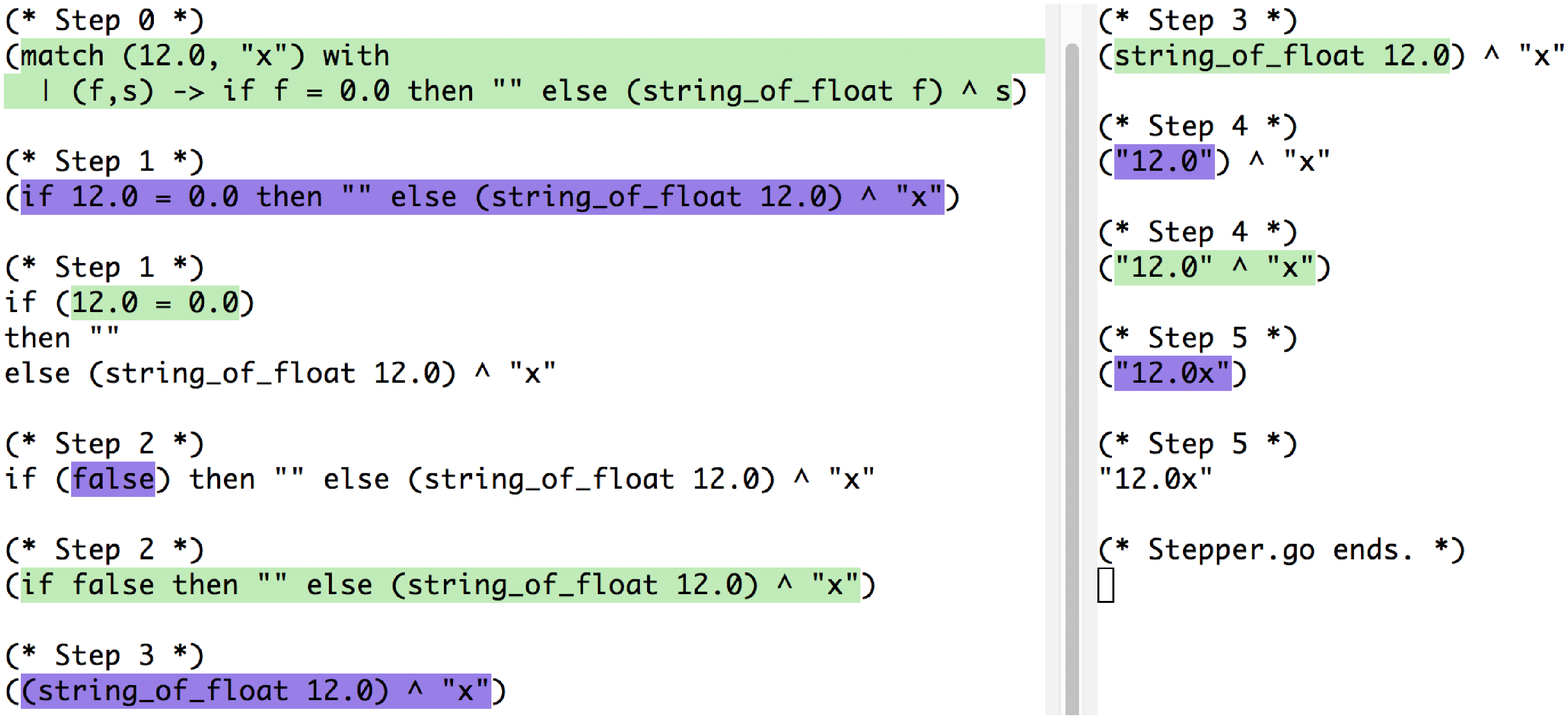}
  \caption{Evaluating programs using the actual stepper}
  \label{figure:ocamlstep}
\end{figure}

In Figure \ref{figure:ocamlstep}, we show a reduction sequence produced by the actual stepping evaluator.  The evaluator supports the following syntactic constructs:

\begin{itemize}
\item integers, floating point numbers, booleans, characters, strings
\item lists, tuples, records
\item user-defined datatypes
\item conditionals, let-expressions, recursive functions, pattern-matching
\item exception handling operators
\item printing functions and sequential execution
\item the List module, user-defined modules
\item references, arrays
\end{itemize}


\begin{figure}
  \includegraphics[width=10cm]{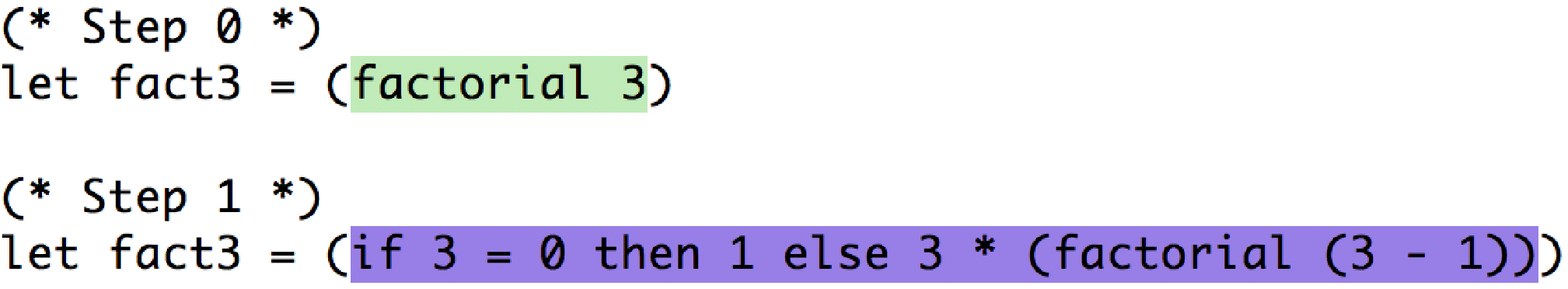}
  \includegraphics[width=4.3cm]{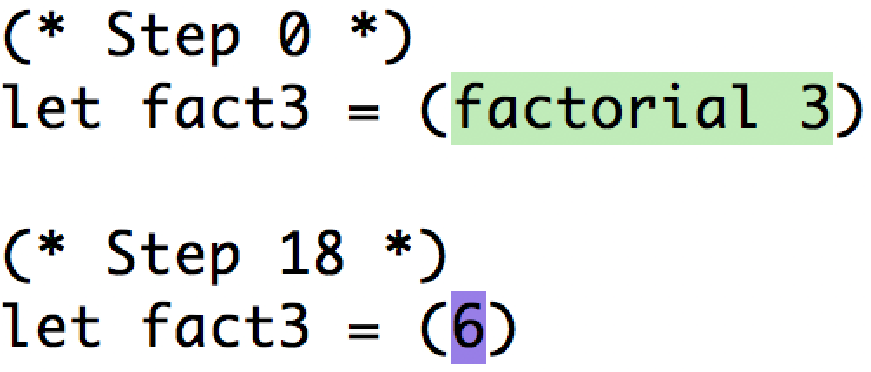}
  \caption{Skipping evaluation of the factorial function}
  \label{figure:factskip}
\end{figure}

To allow the user to adjust granularity of steps, we provide an option for skipping the evaluation of the current function application.  Let us look at Figure \ref{figure:factskip}, which shows skipping of the factorial function.  By pressing the ``skip'' button, we can directly go from the program on the left to the one on the right, without seeing the intermediate steps that appear during the evaluation of the function's body.  This feature helps us focus on the steps we are interested in, allowing us to grasp the overall flow of the execution.

The skipping feature requires some modifications to the \texttt{eval} function (Figure \ref{figure:skipapp}).  The idea is to sandwich the steps within an application between two strings: \texttt{(* Application n start *)} and \texttt{(* Application n end *)}.  Here, \texttt{n} tells us at which step we have entered the application.  These strings are printed using the \texttt{apply\_start} and \texttt{apply\_end} functions, and help the Emacs Lisp program to hide unnecessary steps.  We show an example output sequence in Figure \ref{figure:skipping}.

\begin{figure}
\begin{alltt}
let rec eval expr ctxt = match expr with
    ...
  | App (e1, e2) ->
    begin
      let v2 = eval e2 (add ctxt (CAppR e1)) in
      let v1 = eval e1 (add ctxt (CAppL v2)) in
      match v1 with
      | Lam (x, e) ->
        let e' = subst e x v2 in
        \colorbox{lightgray}{let apply_num = apply_start () in}                (* output start mark *)
        memo (App (v1, v2)) e' ctxt;
        let v = eval e' ctxt in
        \colorbox{lightgray}{apply_end apply_num;}                               (* output end mark *)
        v
      | _ -> failwith "not a function"
    end
  | ...
\end{alltt}
\caption{Skipping application}
\label{figure:skipapp}
\end{figure}

\begin{figure}
\texttt{(* Step 0 *) (f 4) + \colorbox{lightgreen}{10 * 100}\\
(* Step 1 *) (f 4) + \colorbox{purple}{1000}\\
(* Application 1 start *)\\
(* Step 1 *) \colorbox{lightgreen}{f 4} + 1000\\
(* Step 2 *) \colorbox{purple}{(4 * 2) - 1} + 1000\\
(* Step 2 *) \colorbox{lightgreen}{(4 * 2} - 1) + 1000\\
(* Step 3 *) \colorbox{purple}{(8} - 1) + 1000\\
(* Step 3 *) \colorbox{lightgreen}{8 - 1} + 1000\\
(* Step 4 *) \colorbox{purple}{7} + 1000\\
(* Application 1 end *)\\
(* Step 4 *) \colorbox{lightgreen}{7 + 1000}\\
(* Step 5 *) \colorbox{purple}{1007}}
\caption{Stepping application}
\label{figure:skipping}
\end{figure}

\section{Stepping OCaml in the Classroom}
\label{sec:exp}
Since 2016, we have been using (earlier versions of) our stepper in an
introductory OCaml course called ``Functional Programming'',
taught by the third author at Ochanomizu University.

\subsection{The OCaml Course}
The ``Functional Programming'' course teaches how to program with functions and types, covering
basic topics such as recursion, datatypes, effects, and modules.
The course consists of 15, weekly lab sessions, and each session
 consists of 90 minutes lab-style class per week.
(Many students remain in the lab after 90 minutes up until around 150
minutes.)
Throughout the course, students build a program that searches for
the shortest path based on Dijkstra's algorithm.
The participants of the course are second-year undergraduate students
majoring in computer science (around 40 students each year).
All students enter this course after a CS 1 course in the C
programming language.

The course is taught in a ``flipped classroom'' style.
Before every meeting, students are asked to study assigned readings
and videos prepared by the instructor and answer simple quizzes.
In the classroom, they practice the newly covered topics
through exercises, with assistance of the instructor as well as five
to six teaching assistants (including the first and second authors).  

The exercises include simple practice problems and report problems.
The former are for confirming students' understanding of the
topics and are expected to be completed within a class.
The latter problems (for credit) are due in one week.
Whenever a student executes a program, by either step execution or
standard execution, the program as well as its execution log (syntax
errors, type errors, or the result of execution) are recorded.

For most of the problems (up to the 12th week), we provide a check system
where students can submit their solutions to see whether they pass the
given tests.
To earn points for report problems, students are required to have their
programs pass the check system.

\subsection{The Uses of the Stepper}
We introduced the stepper into Functional Programming in 2016.
Although the steppers used in the past had almost the same user-interface as the one shown in Figure \ref{figure:ocamlfac}, they differ in the following ways:
\begin{description}
\item[2016] This first version supported function definitions,
conditionals, records, lists, and recursion.  However, there were
various operations that were not supported.
As such, the usability of the stepper was low.
Moreover, when the instructor introduced the stepper to students, he
only mildly encouraged to use it.
Although we do not know how much the stepper was used in 2016 since we
did not log the execution of stepper, we expect it was used
only rarely in the first few weeks of the course.
\item[2017] Based on the lessons from the previous year, the second
version supported most operations used in the first six weeks.
The instructor introduced the stepper up front at the first class and
showed how to use the stepper with various examples in the subsequent
classes.
\item[2018] The third version supported almost all the constructs
needed for the course, including modules, exception handling,
sequential execution, printing, references, and arrays.
It also supported skipping of function application.
The instructor introduced the stepper as though the stepper was the
only way to execute OCaml programs, encouraging the uses of the
stepper.
(Students gradually realized that they could execute a program in the
interpreter in a few weeks.)
\end{description}

\begin{figure}
  \begin{center}
    \includegraphics[width=15cm]{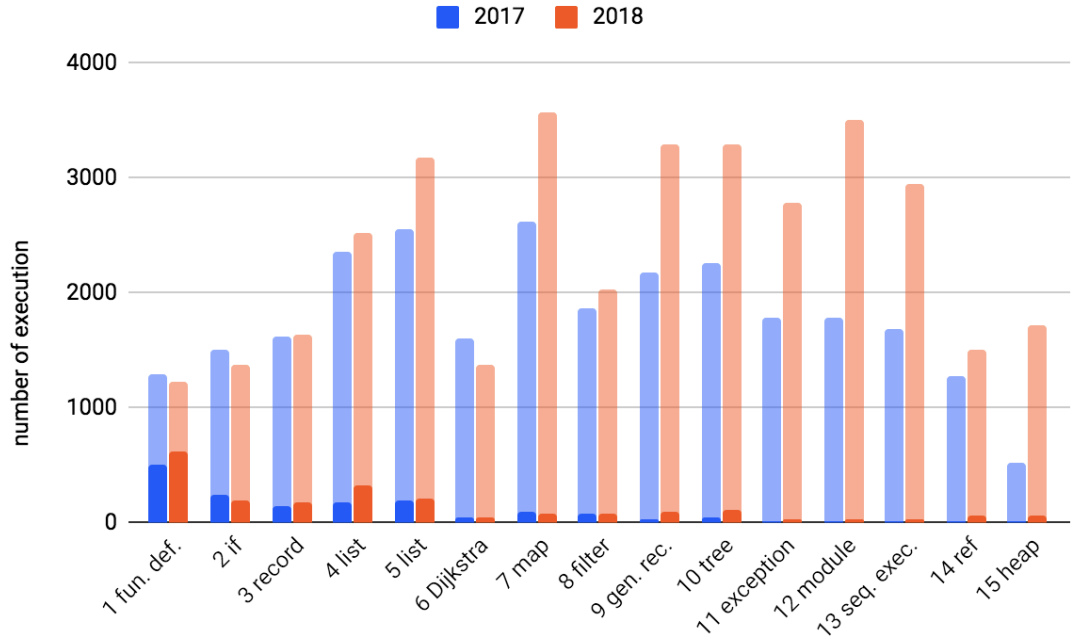}
    \caption{Frequency of standard execution (light-colored) and step execution (dark-colored) in each week in 2017 and 2018. The stepper was not used at all toward the end of the course in 2017, but it was used in some degree in 2018.}
    \label{figure:allExecution}
  \end{center}
\end{figure}

\begin{figure}[!t]
  \begin{center}
    \includegraphics[width=15cm]{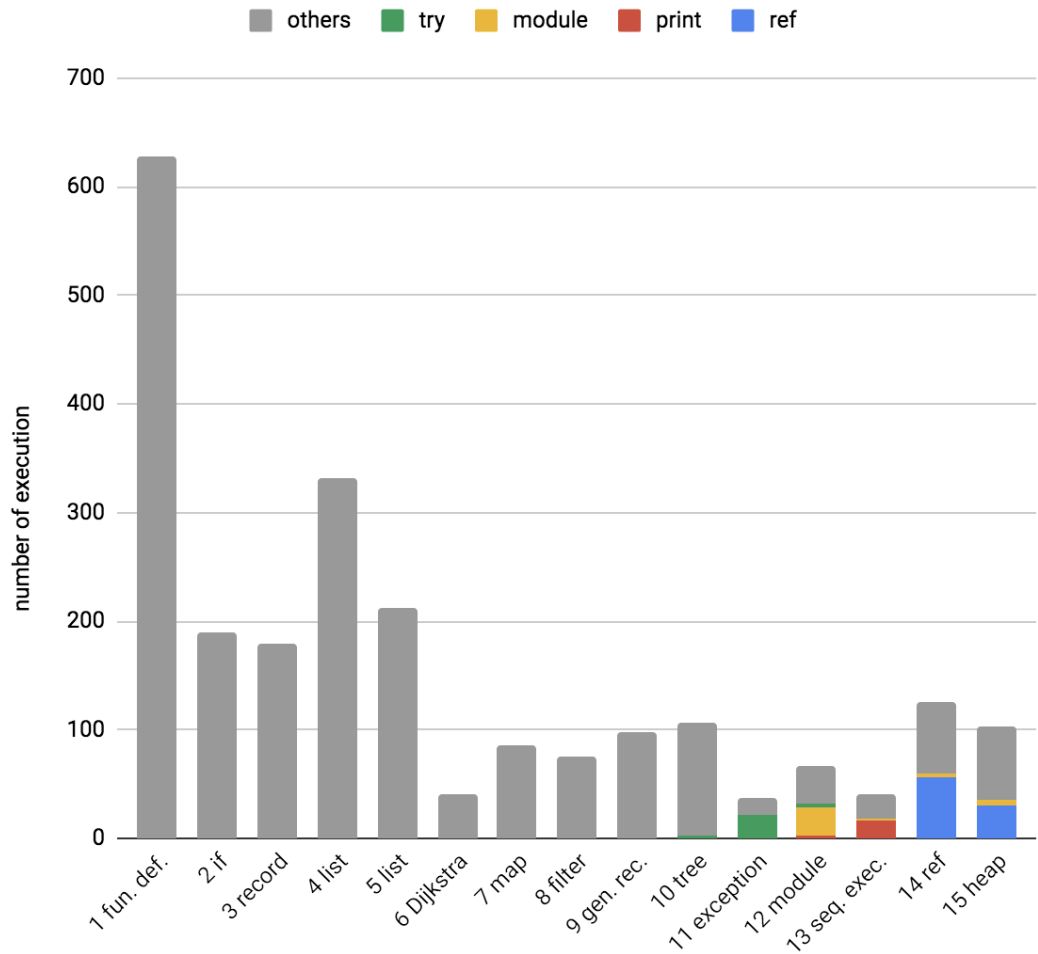}
    \caption{Number of times the stepper was used to evaluate a program with ``try'', ``module'', ``print'' or ``ref'' in 2018.}
    \label{figure:stepExecution}
  \end{center}
\end{figure}

Figure \ref{figure:allExecution} shows how many times students used the stepper among all the executions including the ones that ended up in an error.
In both 2017 and 2018, the stepper was used quite often until week 5.
This is partly because
we encouraged students to use the stepper when they had
trouble finding bugs and understanding recursion.
After week 5, the number decreases, because students started using an
interpreter, too, as programs became larger.

In 2017, the number of stepper uses decreases toward the end of the
course.
In contrast, in 2018, certain number of stepper uses is observed,
thanks to the support of exception handling, modules, and references.
Figure \ref{figure:stepExecution} shows the number of execution of programs using these features during step-execution in 2018. 
From the figure, we can see that there is a demand for step
execution of advanced constructs such as exception handling and
modules.

The exact numbers of execution are available in Table \ref{TableUsage} in the Appendix.

\subsection{Effects of Stepper}
It is not easy to see the effect of a tool like a stepper on
the learning of students.
In the case of improving error messages of a compiler, for example,
one can classify various errors and see how many of them are covered
by the improved error messages objectively.
For the stepper, it is unclear how to show such numeric data.

\begin{figure}
  \begin{center}
    \includegraphics[width=15cm]{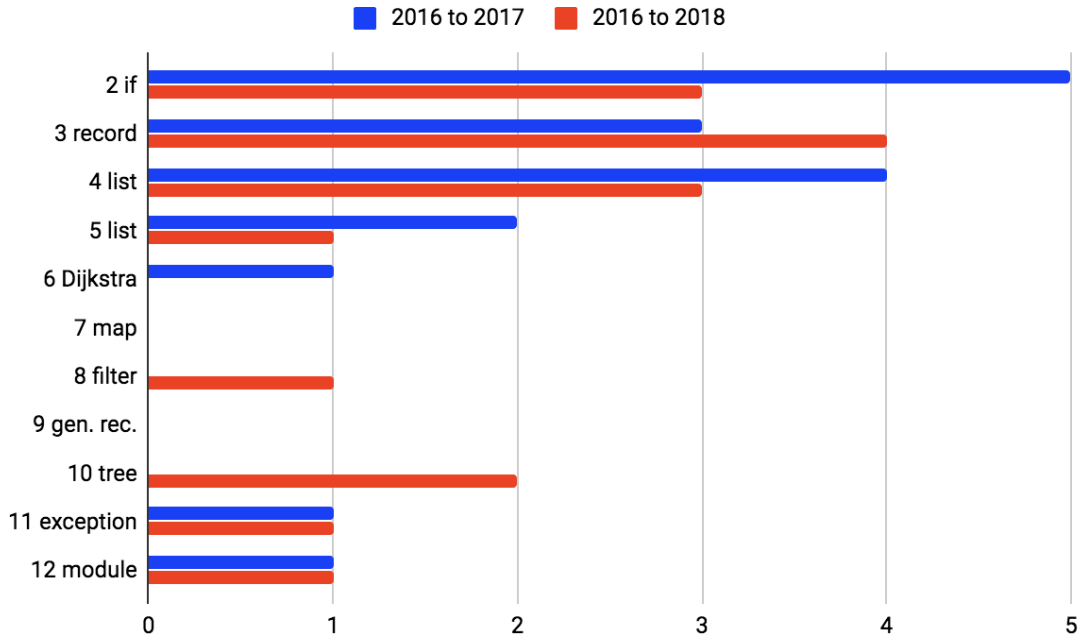}
    \caption{Number of questions where students arrived at a correct answer significantly faster in 2017 and 2018 than 2016.}
    \label{figure:p}
  \end{center}
\end{figure}

As an attempt to measure the effect of the stepper, we examined how
long students took to submit correct solutions to the check system.
Among all the submitted correct solutions, we gathered the (wall-clock) times of
submissions recorded in the check system that are within 100 minutes
from the beginning of the class and compared the average times among
2016, 2017, and 2018.

Figure \ref{figure:p} shows the number of questions for which students submitted a correct answer within significantly shorter time in 2017 and 2018 compared to 2016.  The data are based on one-sided t-testing with p-value $< 0.05$; we refer the reader to Table \ref{TableTTest} of the appendix for details.
Note that we did not include week 1 because we had a special pre-test in the first lecture in 2018.

From the figure, we can see improvement of submission times after the (real) introduction of the stepper, especially in earlier problems.
However, there is an exception: for one problem in the 6th week, correct submissions come significantly later in 2018 than in 2016.
The problem simply asks students to write a recursive function that adds 1 to each
element of a given list.
We do not know why it took so long in 2018.
The result of t-testing all the problems together is t(1778) = 2.819 (p=0.002) in 2017 and t(2111) = 2.592 (p=0.005) in 2018.

We also compared the average times between 2017 and 2018.
For earlier weeks (up to week 5), submissions in 2017 were
significantly earlier, while for later weeks,
there were no significant difference
(except for two problems where the average
times for 2018 were earlier).
Putting all the problems together, the two were not significantly
different with t(1953)=0.455 (p=0.324).

\paragraph{Threats to validity.}
It is possible that the results of our experiment were affected by the enrolled students in each year (there was no over-lapping).
In all the three years, 
the instructor started the class with some introductory comments that
vary in length.
Although the instructor made similar comments in each year, they were
not exactly the same, which could have affected.

\subsection{Students' Evaluation}

\begin{table}
  \begin{center}
  \begin{tabular}{|l|c|r|}
    \hline
    & score & \# of students \\ \hline
    Using the stepper, I could almost always understand & & \\
    the behavior of programs or the cause of errors. & 4 & 3 \\ \hline
    Using the stepper, I could often solve problems at hand. & 3 & 8\\ \hline
    Using the stepper, I could sometimes solve problems at hand. & 2 & 25\\ \hline
    I could rarely find new things using the stepper. & 1 & 2\\ \hline
    The stepper was useless.  I did not use the stepper. & 0 & 0\\ \hline
  \end{tabular}
  \end{center}
  \caption{Students' scoring of the stepper in 2018.
  38 students out of 42 answered.
  The average is 2.3.}
  \label{TableScore}
\end{table}

At the end of the semester of 2018, we asked the students to share
their thoughts on the stepper.
We received responses from 38 out of 42 students.
We first asked whether the stepper was useful on the 0 to 4 point scale.
The results, which we present in Table \ref{TableScore}, suggest
that the stepper is not a silver bullet that is useful
for all the time.
However, most students could solve the problems at hand sometimes
using the stepper.
We are encouraged to see some students choose ``the stepper was almost always useful''.

We next asked students to write when the stepper was useful (if any),
such as when they found their misunderstanding, or when they could
deepen their understanding.
We summarize the answers in two categories.

\paragraph{Understanding of the behavior of programs.}
Seven students answered they could deepen their understanding of
the behavior of programs.
In particular, five students among them wrote explicitly that the
stepper helped them figure out how functions consume recursive data.
We imagine it was particularly instructive to see how a recursive 
function definition is unfolded in nesting application.

Other students answered that they could observe the behavior of
programs in general.
They found that arguments of a function are evaluated before the
function call, and that the elements of a list are evaluated one by one.
Among them, one student observed the right-to-left execution
employed in OCaml.
Previously, such subtle behavior was taught only in passing without
much emphasis.

\paragraph{Debugging.}
Many students found the stepper useful for debugging.
Sixteen students answered they could find what was wrong when their
program did not pass test cases.
By observing each step of execution, they could identify when the
program behaved differently from their expectation.
This is an important step toward debugging in general.
Because printing (and side effects) is handled at the end of the
course, the only debugging method for students had been unit testing:
they checked whether all the component functions worked as expected.
With the stepper, they can simply observe execution of the program and
see when it goes wrong.

Three students found the stepper useful to understand why their
program did not terminate.
Without printing, it is not easy for students to identify the cause of
infinite loops.
Using the stepper, one of the students could not only observe
the infinite loop, but also see how far her program went well
and when it went wrong.

\section{Related Work}
\label{sec:related}

Clements et al.\ \cite{clements01} describe the implementation of the DrRacket stepper.  They point out that when building a stepper, we have to make sure that (i) it displays every element of the reduction sequence in the correct order; and (ii) it has access to information necessary for reconstructing the whole program, namely evaluation contexts.  Based on this idea, they define the stepper as a composition of the following three functions:

\begin{itemize}
\item A breakpoint-inserting function, which places a breakpoint to every piece of a program where reduction takes place
\item An annotating function, which decorates the user program so that it manipulates evaluation contexts appropriately
\item A reconstructing function, which builds the whole program using the context accumulated in the stack
\end{itemize}

As can be inferred, where they insert breakpoints exactly corresponds to where we insert the \texttt{memo} function.  Contexts are handled via two primitives of Racket: \wcm\ (``with-continuation-mark'') and \ccm\ (``current-continuation-marks'').  The former can be understood as extending the context list of our stepper, whereas the latter collects all the context frames on the stack.  The reconstructing function plays the same role as our \texttt{plug} function.  

PLT Redex \cite{felleisen09} is a domain specific language for formalizing operational semantics.  The language provides facilities for defining grammars and reduction rules, and inherits the algebraic stepper from the DrRacket environment \cite{klein12}.  In addition to these, Redex has the ability of generating \emph{reduction graphs} of programs.  A reduction graph is essentially a graphical version of a stepper's output, where the elements of a reduction sequence are connected with arrows.  Reduction graphs are more informative in that each arrow is annotated with the reduction rule used in that step.

Tunnel Wilson et al.\ \cite{tunnell18} investigate students' understanding of functions and recursion via tracing activities.  Instead of adopting the traditional tracing method that uses stacks, they take a substitution-based approach, where students rewrite programs using a set of reduction rules.  The rewriting activity can be viewed as writing down the output of a stepper by hand, although their purpose was not to assist debugging.

\section{Conclusion}
\label{sec:conc}

In this paper, we presented an OCaml stepper that supports advanced
features including exception handling.
The main idea is to keep track of two levels of evaluation contexts,
which we use to reconstruct pre- and post-reduction programs.
We also shared students' feedback on our stepper.
Although it is not easy to measure the effectiveness of the stepper in
education, we reported that the stepper is used by many students with
positive reaction.

The ``Functional Programming'' course is taught every year,
which means we are constantly having new users of our stepper.
As future work, we intend to find other ways to assess the stepper,
and further evaluate its impact on students.
We believe that this would provide new insight into
the development and assessment of pedagogical tool in general.



\section*{Acknowledgments}
We are grateful to anonymous reviewers for constructive comments and
criticisms.
This work was partly supported by JSPS KAKENHI under Grant
No.~15K00090.

\bibliographystyle{eptcs}
\bibliography{stepper}

\begin{thebibliography}{1}
\providecommand{\bibitemdeclare}[2]{}
\providecommand{\surnamestart}{}
\providecommand{\surnameend}{}
\providecommand{\urlprefix}{Available at }
\providecommand{\url}[1]{\texttt{#1}}
\providecommand{\href}[2]{\texttt{#2}}
\providecommand{\urlalt}[2]{\href{#1}{#2}}
\providecommand{\doi}[1]{doi:\urlalt{http://dx.doi.org/#1}{#1}}
\providecommand{\bibinfo}[2]{#2}

\bibitemdeclare{inproceedings}{clements01}
\bibitem{clements01}
\bibinfo{author}{John \surnamestart Clements\surnameend},
  \bibinfo{author}{Matthew \surnamestart Flatt\surnameend} \&
  \bibinfo{author}{Matthias \surnamestart Felleisen\surnameend}
  (\bibinfo{year}{2001}): \emph{\bibinfo{title}{Modeling an algebraic
  stepper}}.
\newblock In: {\sl \bibinfo{booktitle}{European symposium on programming}},
  \bibinfo{organization}{Springer}, pp. \bibinfo{pages}{320--334},
  \doi{10.1007/3-540-45309-1_21}.

\bibitemdeclare{inproceedings}{DF1990}
\bibitem{DF1990}
\bibinfo{author}{Olivier \surnamestart Danvy\surnameend} \&
  \bibinfo{author}{Andrzej \surnamestart Filinski\surnameend}
  (\bibinfo{year}{1990}): \emph{\bibinfo{title}{Abstracting Control}}.
\newblock In: {\sl \bibinfo{booktitle}{Proceedings of the 1990 ACM Conference
  on Lisp and Functional Programming}}, \bibinfo{series}{LFP '90}, pp.
  \bibinfo{pages}{151--160}, \doi{10.1145/91556.91622}.

\bibitemdeclare{book}{felleisen09}
\bibitem{felleisen09}
\bibinfo{author}{Matthias \surnamestart Felleisen\surnameend},
  \bibinfo{author}{Robert~Bruce \surnamestart Findler\surnameend} \&
  \bibinfo{author}{Matthew \surnamestart Flatt\surnameend}
  (\bibinfo{year}{2009}): \emph{\bibinfo{title}{Semantics engineering with PLT
  Redex}}.
\newblock \bibinfo{publisher}{MIT Press}.

\bibitemdeclare{inproceedings}{FF1986}
\bibitem{FF1986}
\bibinfo{author}{Matthias \surnamestart Felleisen\surnameend} \&
  \bibinfo{author}{Daniel~P. \surnamestart Friedman\surnameend}
  (\bibinfo{year}{1986}): \emph{\bibinfo{title}{Control operators, the
  SECD-machine, and the $\lambda$-calculus}}.
\newblock In \bibinfo{editor}{M.~\surnamestart Wirsing\surnameend}, editor:
  {\sl \bibinfo{booktitle}{Formal Description of Programming Concepts III}},
  \bibinfo{publisher}{Elsevier}, pp. \bibinfo{pages}{193--219}.
\newblock \urlprefix\url{https://cs.indiana.edu/ftp/techreports/TR197.pdf}.

\bibitemdeclare{inproceedings}{klein12}
\bibitem{klein12}
\bibinfo{author}{Casey \surnamestart Klein\surnameend}, \bibinfo{author}{John
  \surnamestart Clements\surnameend}, \bibinfo{author}{Christos \surnamestart
  Dimoulas\surnameend}, \bibinfo{author}{Carl \surnamestart
  Eastlund\surnameend}, \bibinfo{author}{Matthias \surnamestart
  Felleisen\surnameend}, \bibinfo{author}{Matthew \surnamestart
  Flatt\surnameend}, \bibinfo{author}{Jay~A. \surnamestart
  McCarthy\surnameend}, \bibinfo{author}{Jon \surnamestart Rafkind\surnameend},
  \bibinfo{author}{Sam \surnamestart Tobin-Hochstadt\surnameend} \&
  \bibinfo{author}{Robert~Bruce \surnamestart Findler\surnameend}
  (\bibinfo{year}{2012}): \emph{\bibinfo{title}{Run Your Research: On the
  Effectiveness of Lightweight Mechanization}}.
\newblock In: {\sl \bibinfo{booktitle}{Proceedings of the 39th Annual ACM
  SIGPLAN-SIGACT Symposium on Principles of Programming Languages}},
  \bibinfo{series}{POPL '12}, \bibinfo{publisher}{ACM}, \bibinfo{address}{New
  York, NY, USA}, pp. \bibinfo{pages}{285--296}, \doi{10.1145/2103656.2103691}.

\bibitemdeclare{inproceedings}{tunnell18}
\bibitem{tunnell18}
\bibinfo{author}{Preston \surnamestart Tunnell~Wilson\surnameend},
  \bibinfo{author}{Kathi \surnamestart Fisler\surnameend} \&
  \bibinfo{author}{Shriram \surnamestart Krishnamurthi\surnameend}
  (\bibinfo{year}{2018}): \emph{\bibinfo{title}{Evaluating the Tracing of
  Recursion in the Substitution Notional Machine}}.
\newblock In: {\sl \bibinfo{booktitle}{Proceedings of the 49th ACM Technical
  Symposium on Computer Science Education}}, \bibinfo{series}{SIGCSE '18},
  \bibinfo{organization}{ACM}, pp. \bibinfo{pages}{1023--1028},
  \doi{10.1145/3159450.3159479}.

\end{thebibliography}

\appendix
\section{Appendix}

\begin{table}[!b]
\begin{center}
  \begin{tabular}{|c||c|c|c|c|c|c||c|c|c|c|c|c||l|}
    \hline
    & \multicolumn{6}{|c||}{2017} & \multicolumn{6}{|c||}{2018} & \\ \cline{2-13}
    \hspace{-1mm}week\hspace{-1mm} & all & step. & try & \hspace{-1mm}mod.\hspace{-1mm} & \hspace{-1mm}print\hspace{-1mm} & ref
    & all & step. & try & \hspace{-1mm}mod.\hspace{-1mm} & \hspace{-1mm}print\hspace{-1mm} & ref & contents\\ \hline
    1 & 1293 & 504 & 0 & 0 & 0 & 0 & 1233 & 627 & 0 & 0 & 0 & 0 & fun.\ def.\\ \hline
    2 & 1511 & 235 & 0 & 0 & 0 & 0 & 1375 & 189 & 0 & 0 & 0 & 0 & if\\ \hline
    3 & 1618 & 144 & 0 & 0 & 0 & 0 & 1641 & 179 & 0 & 0 & 0 & 0 & record\\ \hline
    4 & 2364 & 169 & 0 & 0 & 0 & 0 & 2517 & 332 & 0 & 0 & 0 & 0 & list\\ \hline
    5 & 2556 & 193 & 0 & 0 & 0 & 0 & 3173 & 213 & 0 & 0 & 0 & 0 & list 2\\ \hline
    6 & 1596 & 43 & 0 & 0 & 0 & 0 & 1369 & 41 & 0 & 0 & 0 & 0 & Dijkstra\\ \hline
    7 & 2621 & 92 & 0 & 0 & 0 & 0 & 3570 & 86 & 0 & 0 & 0 & 0 & map\\ \hline
    8 & 1874 & 81 & 0 & 0 & 0 & 0 & 2028 & 75 & 0 & 0 & 0 & 0 & filter\\ \hline
    9 & 2184 & 34 & 0 & 0 & 0 & 0 & 3300 & 98 & 0 & 0 & 0 & 0 & gen.\ rec.\\ \hline
    10 & 2254 & 48 & 0 & 0 & 0 & 0 & 3298 & 106 & 3 & 0 & 0 & 0 & tree\\ \hline
    11 & 1783 & 20 & 10 & 0 & 0 & 0 & 2790 & 37 & 22 & 0 & 0 & 0 & exception\\ \hline
    12 & 1785 & 12 & 8 & 4 & 0 & 0 & 3501 & 37 & 3 & 26 & 3 & 0 & module\\ \hline
    13 & 1678 & 10 & 0 & 7 & 2 & 0 & 2943 & 22 & 0 & 3 & 16 & 0 & seq.\ exec.\\ \hline
    14 & 1280 & 11 & 0 & 0 & 0 & 1 & 1511 & 65 & 0 & 4 & 0 & 56 & ref\\ \hline
    15 & 517 & 6 & 0 & 0 & 0 & 0 & 1717 & 68 & 0 & 5 & 0 & 30 & heap\\ \hline
  \end{tabular}
\end{center}
  \caption{Number of uses of the stepper (step.) among all the
  executions (all) in 2017 and 2018.  The columns try, mod., print,
  and ref represent number of uses of the stepper for programs that
  contain exception handling, modules, printing (and sequential
  execution), and references (including arrays), respectively.
  The rightmost column shows representative topics handled in the
  week.}
  \label{TableUsage}
\end{table}

\begin{table}
\begin{center}  
\small
  \begin{tabular}{|l||c|c|c||c|c|c||l|}
    \hline
    week.
    & \multicolumn{3}{|c||}{2016 to 2017}
    & \multicolumn{3}{|c||}{2016 to 2018}
    &
    \\ \cline{2-7}
    problem & t & p & +/- & t & p & +/- & contents
    \\ \hline
    2.r1
    & t(55)=2.098 & p=\textcolor{red}{0.020} & dec
    & t(60)=0.635 & p=0.264 & dec
    & if\\ \cline{1-7}
    2.r2
    & t(56)=2.364 & p=\textcolor{red}{0.011} & dec
    & t(57)=1.831 & p=\textcolor{red}{0.036} & dec
    & \\ \cline{1-7}
    2.r3
    & t(54)=1.896 & p=\textcolor{red}{0.032} & dec
    & t(59)=0.751 & p=0.228 & dec
    & \\ \cline{1-7}
    2.1
    & t(66)=3.006 & p=\textcolor{red}{0.002} & dec
    & t(74)=3.372 & p=\textcolor{red}{0.001} & dec
    & \\ \cline{1-7}
    2.2
    & t(56)=3.672 & p=\textcolor{red}{0.000} & dec
    & t(62)=3.036 & p=\textcolor{red}{0.002} & dec
    & \\ \hline
    3.r1
    & t(52)=3.222 & p=\textcolor{red}{0.001} & dec
    & t(61)=2.936 & p=\textcolor{red}{0.002} & dec
    & record\\ \cline{1-7}
    3.r2
    & t(42)=2.339 & p=\textcolor{red}{0.012} & dec
    & t(56)=3.467 & p=\textcolor{red}{0.001} & dec
    & \\ \cline{1-7}
    3.r3
    & t(41)=1.373 & p=0.089 & dec
    & t(51)=2.688 & p=\textcolor{red}{0.005} & dec
    & \\ \cline{1-7}
    3.1
    & t(28)=5.610 & p=\textcolor{red}{0.000} & dec
    & t(38)=2.753 & p=\textcolor{red}{0.004} & dec
    & \\ \cline{1-7}
    3.2
    & t(17)=1.655 & p=0.058 & dec
    & t(27)=0.105 & p=0.459 & dec
    & \\ \cline{1-7}
    3.3
    & t(16)=1.546 & p=0.071 & dec
    & t(13)=0.603 & p=0.279 & dec
    & \\ \hline
    4.r1
    & t(47)=2.088 & p=\textcolor{red}{0.021} & dec
    & t(61)=2.446 & p=\textcolor{red}{0.009} & dec
    & list\\ \cline{1-7}
    4.r2
    & t(48)=1.909 & p=\textcolor{red}{0.031} & dec
    & t(60)=2.267 & p=\textcolor{red}{0.014} & dec
    & \\ \cline{1-7}
    4.1
    & t(51)=2.134 & p=\textcolor{red}{0.019} & dec
    & t(60)=2.473 & p=\textcolor{red}{0.008} & dec
    & \\ \cline{1-7}
    4.2
    & t(18)=3.033 & p=\textcolor{red}{0.004} & dec
    & t(20)=0.489 & p=0.315 & dec
    & \\ \hline
    5.r1
    & t(42)=1.037 & p=0.153 & dec
    & t(55)=0.257 & p=0.399 & inc
    & list 2\\ \cline{1-7}
    5.1
    & t(49)=1.592 & p=0.059 & dec
    & t(61)=0.904 & p=0.185 & dec
    & \\ \cline{1-7}
    5.2
    & t(55)=4.138 & p=\textcolor{red}{0.000} & dec
    & t(62)=1.631 & p=0.054 & dec
    & \\ \cline{1-7}
    5.3
    & t(47)=3.305 & p=\textcolor{red}{0.001} & dec
    & t(50)=1.940 & p=\textcolor{red}{0.029} & dec
    & \\ \hline
    6.r1
    & t(30)=0.322 & p=0.375 & inc
    & t(51)=2.011 & p=\textcolor{blue}{0.025} & inc
    & Dijkstra's algorithm\\ \cline{1-7}
    6.1
    & t(41)=1.678 & p=0.050 & dec
    & t(61)=1.155 & p=0.126 & dec
    & \\ \cline{1-7}
    6.2
    & t(45)=1.415 & p=0.082 & dec
    & t(62)=0.976 & p=0.166 & dec
    & \\ \cline{1-7}
    6.3
    & t(34)=2.296 & p=\textcolor{red}{0.014} & dec
    & t(42)=0.548 & p=0.293 & dec
    & \\ \hline
    7.r1
    & t(42)=0.462 & p=0.323 & inc
    & t(56)=0.314 & p=0.377 & dec
    & map\\ \cline{1-7}
    7.r2
    & t(41)=0.286 & p=0.388 & inc
    & t(54)=1.181 & p=0.121 & dec
    & \\ \cline{1-7}
    7.r3
    & t(40)=0.677 & p=0.251 & inc
    & t(51)=1.492 & p=0.071 & dec
    & \\ \cline{1-7}
    7.1
    & t(21)=0.965 & p=0.173 & dec
    & t(20)=0.372 & p=0.357 & dec
    & \\ \cline{1-7}
    7.2
    & t(12)=0.380 & p=0.355 & inc
    & \, t(7)=0.686 & p=0.258 & dec
    & \\ \hline
    8.r1
    & t(46)=1.162 & p=0.126 & inc
    & t(58)=2.694 & p=\textcolor{red}{0.005} & dec
    & filter\\ \cline{1-7}
    8.1
    & t(16)=0.844 & p=0.205 & dec
    & t(22)=0.841 & p=0.205 & dec
    & \\ \hline
    9.r1
    & t(44)=1.294 & p=0.101 & dec
    & t(55)=0.678 & p=0.250 & dec
    & general recursion\\ \hline
    10.r1
    & t(48)=0.312 & p=0.378 & dec
    & t(51)=1.308 & p=0.098 & dec
    & tree\\ \cline{1-7}
    10.r2
    & t(48)=0.457 & p=0.325 & dec
    & t(50)=1.760 & p=\textcolor{red}{0.042} & dec
    & \\ \cline{1-7}
    10.1
    & t(33)=0.976 & p=0.168 & dec
    & t(38)=0.845 & p=0.202 & dec
    & \\ \cline{1-7}
    10.2
    & t(19)=1.498 & p=0.075 & dec
    & t(19)=0.871 & p=0.197 & dec
    & \\ \cline{1-7}
    10.3
    & t(15)=1.538 & p=0.072 & dec
    & t(12)=2.240 & p=\textcolor{red}{0.022} & dec
    & \\ \hline
    11.r1
    & t(43)=0.272 & p=0.393 & dec
    & t(53)=1.555 & p=0.063 & dec
    & exception\\ \cline{1-7}
    11.r2
    & t(37)=0.454 & p=0.326 & dec
    & t(44)=1.906 & p=\textcolor{red}{0.032} & dec
    & \\ \cline{1-7}
    11.r3
    & t(31)=0.050 & p=0.480 & inc
    & t(40)=1.208 & p=0.117 & dec
    & \\ \cline{1-7}
    11.1
    & t(34)=1.567 & p=0.063 & dec
    & t(46)=1.518 & p=0.068 & dec
    & \\ \cline{1-7}
    11.2
    & t(28)=2.204 & p=\textcolor{red}{0.018} & dec
    & t(36)=1.563 & p=0.063 & dec
    & \\ \cline{1-7}
    11.3
    & t(17)=0.604 & p=0.277 & dec
    & t(21)=0.384 & p=0.352 & dec
    & \\ \hline
    12.r1
    & t(39)=2.229 & p=\textcolor{red}{0.016} & dec
    & t(52)=4.009 & p=\textcolor{red}{0.000} & dec
    & module\\ \hline
    all
    & t(1778)=2.819 & p=\textcolor{red}{0.002} & dec
    & t(2111)=2.592 & p=\textcolor{red}{0.005} & dec
    & \\ \hline
  \end{tabular}
  \end{center}
  \caption{Result of one-sided t-test with p-values comparing the time between the beginning of the class and the moment that students submitted a correct solution.  The column +/- shows whether the average time increased or decreased.  The p-values below $0.05$ are colored.}
  \label{TableTTest}
\end{table}

\end{document}